\documentclass[pra,twocolumn]{revtex4-1}
\usepackage{amsmath,amssymb,amsfonts}
\usepackage{ifsym}
\usepackage{amsthm}

\def\ie{i.~e.~}
\def\eg{e.~g.~}

\newcommand{\st}[1]{\mathbf{#1}}
\newcommand{\set}[1]{\mathsf{#1}}

\newcommand{\Tr}{\operatorname{Tr}}

\newcommand{\map}[1]{\mathcal{#1}}

\usepackage{graphicx}
\usepackage{dcolumn}
\usepackage{bm}

\newtheorem{principle}{Principle}

\newtheorem{prop}{Proposition}

\begin{document}

\preprint{APS/123-QED}

\title{
Measurement sharpness cuts nonlocality and contextuality in every physical theory}

\author{Giulio Chiribella}
\author{Xiao Yuan}
\affiliation{Center for Quantum Information, Institute for Interdisciplinary Information Sciences, Tsinghua University, Beijing, 100084, China}

\date{\today}

\begin{abstract}
Gathering data through measurements is at the basis of  every experimental science.  Ideally, measurements should be repeatable and, when extracting only  coarse-grained data, they should allow the experimenter  to retrieve the finer  details at  a later time. 
However,   in practice most  measurements appear to be noisy.   
 Here we postulate  that, despite the imperfections observed in real life experiments, there exists a fundamental level where    all measurements are  ideal. 
Combined with the requirement that ideal measurements remain so when coarse-grained or applied in parallel on spacelike separated systems, 
our postulate places a powerful constraint on the amount of nonlocality and contextuality that can be found in an arbitrary physical theory, 
bringing down  the violation of Bell and Kocher-Specker inequalities near  to its quantum value. In addition, it provides a new compelling motivation for the  principles  of   Local Orthogonality and Consistent Exclusivity, recently proposed  for the characterization of the  quantum set of  probability distributions.  


\end{abstract}

\maketitle


Nonlocality \cite{EPR,bell,chsh} and contextuality \cite{kochen,mermin93} are among the most striking features of quantum mechanics,   
 in radical conflict with the worldview of classical physics.  
 Still, quantum mechanics is neither the most nonlocal theory one can imagine, nor the most contextual.   For nonlocality,  this observation dates back to the seminal work of Popescu and Rohrlich \cite{pr95}, who showed that relativistic  no-signalling  is compatible with correlations that are much stronger than those allowed by  quantum theory.    
Their work stimulated the question whether other fundamental principles, yet to be discovered, characterize the peculiar set of  correlations observed in the quantum world.  Up to now, several candidates that partly retrieve the set of quantum correlations have been proposed, including Non-Trivial Communication Complexity \cite{Dam05, brassard2006}, No-Advantage in Nonlocal Computation \cite{linden2007}, Information Causality \cite{Paw09}, Macroscopic Locality \cite{Navascues09},  and, lately,  Local Orthogonality (LO) \cite{Fritz12}.        The observation that quantum theory is not maximally contextual made an early appearance in Kochen's and Specker's work \cite{specker1960,kochen}, but it was not until  recently that it caught broad attention in the community  \cite{spekkens2011,cabello2013,cabello2014}, initiating a  search for principles that characterize the quantum set of contextual probability distributions.  On this front, the only principle put forward so far is   Consistent Exclusivity  (CE) \cite{cabello2013,henson2012,acin2012,yan2013}. 

Despite many  successes, a complete characterization of the  quantum set is still extremely challenging \cite{almost}. What makes the problem hard is the fact that---intendedly---the principles considered so far dealt only with   input-output probability distributions, without making any hypothesis on how  these distributions are generated.  
  On the other hand, a physical theory does not provide only probability distributions, but also specifies  rules on how to combine physical systems together, how to measure them, and how to evolve their state in time \cite{coecke2010}.   Considering that fundamental quantum features like no-cloning  and  universal   computation cannot be expressed just in terms of input-output distributions,  it is natural to wonder whether also quantum nonlocality and contextuality could be better understood in a broader framework of general probabilistic theories  (GPTs)  \cite{hardy01,barrett07,chiribella10,barnum11}.    Further motivation to extend the framework  comes from the latest principles in the nonlocality and contextuality camps: LO and CE.  Both principles  refer to a notion of orthogonal events and impose that the sum of the  probabilities of a set of mutually orthogonal events shall not exceed one.    This  is a powerful requirement, which in the case of LO is even capable to rule out non-quantum correlations that are compatible with every bipartite principle \cite{gallego2011}.   But why  should Nature obey such a requirement?  And what does this requirement tell us about the fundamental laws that govern   physical processes?  

Here we tackle the problem of understanding quantum nonlocality and contextuality from a new angle, which focuses  on the fundamental structure of  measurements in arbitrary physical theories.     We  introduce a class of ideal measurements, called sharp, that are repeatable and  cause the minimal amount of disturbance on future observations. 
We postulate that all measurements are  sharp at the fundamental level and we explain  the apparent unsharpness of realistic experiments as due to the interaction with the environment surrounding the measured system.  
Assuming that sharp measurements remain sharp under elementary operations, such as joining  two outcomes together  and applying two measurements in parallel,  we show that  the fundamental sharpness of measurements  implies the validity of CE and LO, thus providing a strong constraint on the set of probability distributions and bringing down the violation of Bell and Kochen-Specker inequalities near to their quantum value.   Our result demonstrates that  principles formulated in the broader framework of  GPTs   can offer an extra power in the characterization of the quantum set and  identifies the fundamental sharpness  of measurements as a candidate principle for  future axiomatizations of quantum theory.

\section{Results}
{\bf Framework.}   
In a general physical theory,  a measurement is described by a collection of events,   each event  labelled by an outcome.   We first consider demolition measurements, which adsorb the measured system.    In this case,  
the measurement events are called \emph{effects} and the measurement is a collection of effects $\{m_x\}_{x\in\set X}$.      
For a system prepared in the state $\rho$, the probability of the outcome $x$ is denoted by $p_x  =  (  m_x|\rho)$.    In quantum theory this is a notation for the Born rule $p_x  =  \Tr  [  m_x  \rho]$, where $\rho$ is a density matrix and $m_x$ is a measurement operator.     In general theories, $(m_x|\rho)$ does not denote a trace of matrices and in fact the actual recipe for computing the probability  $(m_x|\rho)$ is irrelevant  here.
We will often use the notation   $(m_x|$   and $ |\rho  ) $ for effects and states,  respectively.  
It is understood that two different states  give different probabilities for at least one effect, and two different effects take place with different probabilities on at least one state.  

 When two measurements $ \{m_x\}$ and $\{n_y\}$ are performed in parallel on two systems $A$ and $B$, we denote by $m_x \otimes n_y$ the measurement event labelled by the pair of outcomes $x,y$.  Similarly, when  two states of systems $A$ and $B$, say $\alpha$ and $\beta$, are prepared independently,  we denote by $\alpha  \otimes \beta$ the corresponding state of the composite system $AB$.   In quantum theory, this is the ordinary tensor product of operators, but this may not be the case in a general theory and, again,  the actual recipe for computing $\alpha\otimes \beta$  is irrelevant here.       
What is relevant, instead, is that the notation is consistent with the operational notion of performing independent operations on different systems:        If two  systems   are independently prepared in states $\alpha$ and $\beta$ and undergo to independent measurements $ \{m_x\}$ and $\{n_y\}$, we impose that the  probability has the product form $ p_{xy}  =    (  m_x|\alpha) \, (n_y  |  \beta)$.  

The most basic operation one can perform on a measurement is to join  some outcomes together, thus obtaining a new, less informative measurement. This operation, known as \emph{coarse-graining},  is achieved by dividing the outcomes  of the original measurement $\{m_x\}_{x\in\set X}$ into disjoint groups $   \{   \set X_z\}_{z\in\set Z}$, and by identifying outcomes that belong to the same group.  The result of this procedure is a new  measurement $ \{m'_z\}_{  z\in\set Z}$ satisfying  the relation 
$(m'_z  |  \rho)  =    \sum_{x\in\set X_z}  (m_x|\rho)$
for every every $z$ and for every possible state $\rho$.   For brevity,  we write $ m'_z  =  \sum_{x\in\set X_z}   m_x$.   

Coarse-graining allows one  to  express the  principle of  \emph{causality},   
 which states that the settings of future measurements do not influence the outcome probabilities of present experiments  \cite{chiribella10}. Causality is   equivalent to the requirement that for every system $A$  there exists an effect   $u_A$, called the \emph{unit},    such that 
\begin{equation}\label{causal} \sum_{x\in\set X}   m_x  =  u_A 
\end{equation} 
for every measurement   $\{m_x\}_{x\in\set X}$   on $A$.    
  In quantum theory,   $u_A$ is the identity operator on the Hilbert space of the system and Eq. (\ref{causal})  expresses the fact that quantum measurements are resolutions of the identity.   When there is no ambiguity, we drop the subscript from $u_A$.  

Causality has major consequences.    First of all, it implies that the probability distributions generated by local measurements satisfy the no-signalling principle \cite{chiribella10}.  
Moreover,  it allows  to perform adaptive operations:  for example, if $\{m_x\}_{x\in \set X}$ is a measurement on system $A$ and  $\{n^{(x)}_y\}_{y\in\set Y}$ is a measurement on system $B$ for every value of $x$, then causality guarantees that it is possible to choose the measurement on $B$ depending on the outcome on system $A$, \ie that $\{  m_x\otimes n^{(x)}_y\}_{x\in\set X,  y\in\set Y}$ is a legitimate measurement.   Finally, causality allows one to describe non-demolition measurements.  
For a non-demolition measurement  $\{\map M_x\}_{x\in\set X}$, the measurement events  are \emph{transformations}, which turn the initial state of the system, say $\rho$, into a new  unnormalized state $  \map M_x   |\rho)$.
For a system prepared in the state $\rho$,   the probability of the outcome $x$  is   $  p_x = (u|  \map M_x  |\rho)$ and, conditionally on outcome $x$,   the post-measurement state is $ \map M_x   | \rho ) /   (u|\map M_x |  \rho)$.   We  will often refer to  the non-demolition measurements as \emph{instruments}, in analogy with the usage in quantum theory  \cite{Davies76,Ozawa84}.   
Note that, thanks to causality, every instrument   $\{\map M_x\}$ is associated to a unique demolition measurement $\{m_x\}$ via  the relation 
\begin{equation}\label{prob}
(m_x|     =   (u|  \map M_x \qquad \forall  x\in\set X\, .
\end{equation}  
 By definition, $\{m_x\}$ describes the statistics of the instrument:  for every state $\rho$ and for every outcome $x$, one has  $p_x   =   (  u  | \map M_x  |\rho)  \equiv   (m_x|\rho)$.   
  
\medskip   
{\bf Sharp measurements in arbitrary theories.}   In textbook  quantum mechanics,  physical quantities are associated to self-adjoint operators, called observables  \cite{sakurai}.  The values of a quantity are the eigenvalues of the corresponding operator and the probability that a measurement outputs the value  $x$ is given by the Born rule $p_x  =  \Tr [  P_x\rho ]$, where $P_x $ is the projector on the eigenspace for the eigenvalue $x$ and $\rho$ is the density matrix of the system before the measurement.  If the measurement gives the outcome $x$,   then the state after the measurement is  $\rho_x'  = P_x\rho P_x/  \Tr[P_x\rho]$, according to the projection postulate. 
These canonical  measurements, where all the measurement operators are orthogonal projectors,  are called  sharp  \cite{Busch96}. 
 While it is clear that sharp measurements play a  key  role  in quantum theory 
 it is by far less clear how to define them in an arbitrary GPT.     Here we propose a  simple definition based on the notions of repeatability and minimal disturbance.  
 
 Let us start from repeatability.   
An instrument   $\{ \map M_x\}$ is \emph{repeatable} if it gives the same outcome when performed two consecutive times,  namely
\begin{align}\label{rep}    (     m_x|  \map M_x    =  (   m_x|    \qquad \forall x\in\set X  \, ,
\end{align} 
where $\{m_x\}$ is the  measurement of Eq. (\ref{prob}).  
Repeatability poses a fairly weak requirement on  $\{m_x\}$:    
every measurement  that discriminates perfectly among a set of states  $\{\rho_x\}$ can be realized by a repeatable instrument, which consists in measuring $\{m_x\}$ and, if the outcome is $x$, re-preparing the system in state $\rho_x$.  

The second ingredient entering in  our definition of  sharp measurements is  minimal disturbance.     We say that the instrument $ \{\map M_x\}_{x\in\set X}$    does not disturb the  measurement   $\st n  = \{n_y\}_{y\in\set Y}$  if   the former does not  affect the statistics of the latter, namely   
\begin{align}\label{nodist}
(  n_y|  \map M  =  (  n_y|  \qquad \forall   y\in\set Y   \, ,     
\end{align}
where   $(n_y| \map M:  =  \sum_{x\in\set X}   (n_y| \map M_x$.  
 Then, we ask which instruments disturb the smallest possible set of measurements.  
 Clearly, if   $\{\map M_x\}$  does not disturb   $\st n$, then  $\st n$ must be compatible with  the measurement $\st m  =  \{  (u|  \map M_x\}$, in the sense that $\st m$ and $\st n$  can be measured jointly.    Indeed, by measuring $\st n$ after   $ \{    \map M_x  \}$ one obtains   the  probability distribution  $ p_{xy}   =   (   n_y  |\map M_x  |  \rho)$, whose  marginals on $x$ and $y$ are equal to the probability distributions of $\st m$ and $\st n$, respectively.    Read in the contrapositive, this means that if $\st m$ and $\st n$ are incompatible,  the instrument $\{\map M_x\}$ must disturb $\st n$.  
This leads us to the following definition:   an instrument  $\{\map M_x\}$ has \emph{minimal disturbance} if it  disturbs only the measurements that are incompatible with   $\st m  = \{  (u|  \map M_x\}$.  

We define an instrument to be sharp if it  is both repeatable and with minimal disturbance. We say that a measurement is  sharp if  it describes the statistics of  a sharp instrument and we call an effect sharp if it belongs to a sharp measurement.     
 In quantum theory, our definition coincides with the usual one:  one can prove that the only sharp instruments are the L\"uders instruments  \cite{Luders50}, of the form $\map M_x  (\rho)   =   P_x  \rho P_x$ where $\{P_x\}$ is a collection of orthogonal projectors.  Hence, the sharp measurements are projective measurements. 
In addition, we can prove that  when a sharp measurement extracts a coarse-grained information,  the experimenter can still retrieve the finer details at a later time. In fact, this is a necessary and sufficient condition for a measurement to be sharp, as proven in the Methods section.

\medskip 

{\bf  Fundamental sharpness of measurements.}       Sharp measurements are an ideal standard---they are the measurements that generate outcomes in a repeatable way, while at the same time causing the least disturbance on future observations.  Unfortunately though, most measurements  in real life  appear to be noisy and not repeatable.  Hence the natural question:  Is noise  fundamental? Or rather it is contingent to the fact that the experimenter has  incomplete control  on the conditions of the experiment?  
Here we state that noise is not fundamental and only arises from the fact that the realistic measurements do not extract information only from the system, but also  from the surrounding environment:     
\begin{principle}[Fundamental Sharpness of Measurements]\label{ax:sharpness} Every measurement arises from a sharp measurement performed jointly on the system and on the  environment.    
\end{principle}
Precisely, we require that for every measurement   $\st m  =  \{m_x\}_{x\in\set X} $   there exists an environment $E$, a state  $\sigma$  of $E$, and a sharp measurement $\st M  =  \{M_x\}_{x\in\set X} $  on the composite system $SE$  such that, for every state $\rho$ of system $S$, one has  $(  m_x | \rho)      =    \left (  M_x \right|      \rho  \otimes   \sigma)$ for every outcome  $ x\in\set X $.
In quantum theory, this is the content of the celebrated Naimark's theorem  \cite{Helstrom76,Holevo11}.    
This is a deep property,  hinting at the idea there exists a  fundamental level where all measurements  are ideal.

\begin{figure}
\centering
\includegraphics[totalheight=50mm]{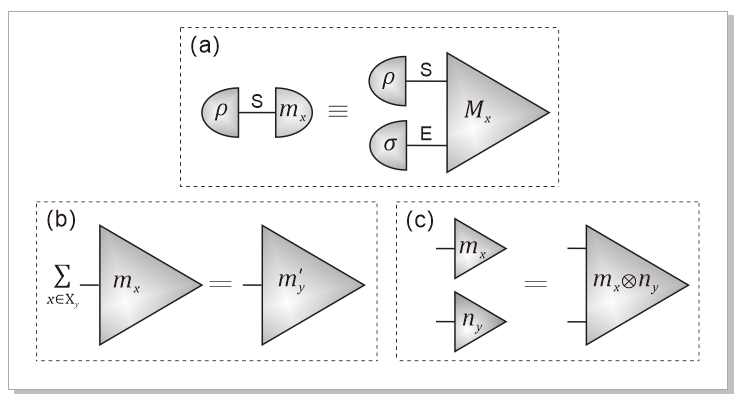}
\caption{{\bf The fundamental structure of sharp  measurements.}    ({\bf a})  Every non-sharp measurement  $\{m_x\}_{x\in\set X}$ (round diagram on the l.h.s.) performed on  system  $S$ in state $\rho$ is equivalent to a sharp measurement   $\{M_x\}_{x\in\set X}$  (triangular diagram on the r.h.s.) performed jointly on the system and on  an environment $E$ in state $\sigma$.   ({\bf b})    Coarse-graining a  sharp measurement $\{ m_x\}_{x\in \set X}$ yields a new  sharp measurement $\{ m'_y \}_{y\in\set Y}$.  ({\bf c})  When two sharp measurements $\{m_x\}$ and $\{m_y\}$  are performed in parallel, they yield a new sharp measurement  $\{m_x \otimes n_y\}$.    
}
\label{fig:sharp}
\end{figure}

Let us push  the idea  further.  If  measurements are sharp at the fundamental level, it is natural to assume that the set of sharp measurements is closed under the basic operation of coarse-graining, which transforms an initial measurement $\st m$ into a new, less informative measurement $\st m'$.  Indeed, since $\st m'$ provides less information than $\st m$, one expects that $\st m'$ should not be less repeatable, nor create more disturbance, than $\st m$.  
 This intuition leads to the following requirement:  
\begin{principle}[Less Information, More Sharpness]\label{ax:coarse}  
If a measurement is less informative than a sharp measurement, then it is sharp.  
\end{principle}
Suppose now that two experimenters,  Alice and Bob, perform two sharp measurements on two systems $A$ and $B$ in their laboratories.    Again, if measurements are sharp at the fundamental level, one expects the result of Alice's and Bob's measurements to be a sharp measurement  on the composite system $A B$.  If this were not the case, it would mean that at the fundamental level some measurements require  nonlocal interactions, even though at the operational level the they appear to be implemented locally by Alice and Bob. 
 We then postulate the following 
\begin{principle}[Locality of Sharp Measurements]\label{ax:loc}  If  two sharp measurements are applied in parallel on  systems $A$ and $B$, then the result is a sharp measurement  on the composite system $AB$.   
\end{principle}    
Principles 1-3 lay down the fundamental structure of sharp measurements, summarized in Fig. \ref{fig:sharp}.   
They  are  satisfied by classical theory and by quantum theory, both on complex and real Hilbert spaces.    In the following we will show that the fundamental structure of sharp measurements has an enormous impact on the amount of nonlocality and contextuality that can be found in a physical theory.   
 

\medskip  

{\bf Derivation of CE.}  
At present, CE  is the only principle known to constrain the amount of contextuality of a generic theory.  
Operationally, the principle can be formulated as follows:   Consider  a collection of sharp measurements $\{\st m^{(x)} , \,    x\in\set X\}$,  each measurement having outcomes in a set $\set Y_x$.   
Suppose  that the possible events have been labelled so that two effects corresponding to the same outcome coincide, \ie   $m_y^{(x)}   \equiv   m_y$, independently  of $x$.   Letting $\set Y  =  \cup_x \set Y_x$ be the set of all  outcomes, one  calls  two distinct outcomes  $y,y'  \in\set Y$ \emph{exclusive} if    there exists a measurement setting $x$   such that  both $y$ and $y'$ belong to  $ \set Y_x$.    We say that   a theory satisfies CE if for every set of mutually exclusive outcomes   $\set E$ and for  every state $\rho$ the probabilities $p_y  =   ( m_y|\rho )$  obey the bound
$\sum_{y\in \set E}   p_y \le 1 \, .$ 

Our first key result is the derivation of  CE.  In fact, we prove a stronger result: We define  two sharp effects  $m$ and $m'$   to be \emph{orthogonal} if they belong to the same  (not necessarily sharp) measurement and   we prove that mutually orthogonal effects can be combined   into a single sharp measurement  (see Methods). Clearly, since mutually exclusive outcomes correspond to mutually orthogonal effects, the existence of a joint measurement  containing the effects  $\{  m_y\}_{y\in\set E}$ implies the bound $\sum_{y\in \set E}   (m_y|  \rho)  \le 1$.    Our result implies that in a theory where measurements are fundamentally sharp  the violation of Kochen-Specker inequalities is upper bounded by the value set by CE  \cite{cabello2014}.  What is remarkable here is that a single requirement on measurements influences directly the strength of contextuality in an arbitrary physical theory.   This situation contrasts with that of the known  axiomatizations of quantum theory  \cite{hardy01,chiribella11,hardy11,masanes,Brukner,masanes12}, where the quantum bounds on contextuality    are retrieved  only indirectly through the derivation of the  Hilbert space framework.  

Our principles do not imply only CE, but also the whole hierarchy of extensions of this principle defined in Ref.   \cite{acin2012}.  The $L$-th level of the hierarchy can be defined by considering independent measurements on $L$ copies of the state $\rho$.  Denoting by  $ \st y  =  (y_1,\dots, y_L)   $  the string of all outcomes,  one says that  two strings $\st y$ and $\st y'$ are exclusive if there exists some $i$ such that $y_i$ and $y_i'$ are exclusive.      A physical theory satisfies the  $L$-th level of the hierarchy if    
 the probabilities  $p_L  (\st y)  =   \prod_{i=1}^L (  m_{y_i}  |  \rho)$ obey the bound   
 $\sum_{\st y  \in  \set E}    p_L (\st y)   \le 1  $
for every set $\set E$ of mutually exclusive strings.  In the Methods section we show that our principles on sharp measurements imply that this bound is satisfied for every possible $L$. 
Again, this fact has major consequences on the amount of contextuality that can be found in a theory satisfying our principles. For example, choosing $L=2$ and invoking a result of Ref. \cite{cabello2013} we have that the structure of sharp measurements implies that the maximum violation of KCBS inequality \cite{klyachko} is exactly equal to the quantum value.

\medskip 

{\bf Derivation of LO.}   In the nonlocality camp, LO  occupies a special position, being up to now the  only known principle   that rules out  non-quantum correlations that are not detected  by any bipartite principle \cite{gallego2011}.    
LO refers to a scenario where $N$  parties perform local measurements on $N$  systems, initially prepared in some  joint state.  The $i$-th party  can choose among different measurement settings in a set $\set X_i$ and her measurements give outcomes in another set $\set Y_i$.   Let $ \st x  =  (x_1,\dots, x_N)  $ be the string of all settings,  $ \st y  =  (y_1,\dots, y_N)   $ be the string of all outcomes, and $\st e$ be the pair $\st e = (\st  x,\st  y)$.  In this context, the pair $\st e = (\st  x,\st  y)$ is called an event and   two events are called \emph{locally orthogonal} iff  there exists a party $i$ such that $x_i =  x_i'$ and $y_i\not=y_i'$.        Setting  $p(\st e)$ to be the conditional probability distribution $   p(\st y| \st x)$,    one says that theory satisfies local orthogonality if all the probability distributions generated by local measurements   obey the bound 
$\sum_{\st e  \in  \set O}  p  ( \st e)   \le 1$
for every  set  $\set O$ of pairwise locally orthogonal events.  

To derive LO, we specify how the probability distribution $p(\st y|\st x)$ is generated:  In the most general scenario,  the $N$ parties share a state $\rho$ and that, for setting $x_i$,  party $i$ performs a measurement  $\st m^{(i,x_i)}$.  Denoting the product effects as $P^{(\st x)}_{\st y}  : =  \bigotimes_{i=1}^N  m^{(i,x_i)}_{y_i} $,  the probability distribution of the outcomes is given by  $  p(\st e)  : =  \left(    P^{(\st x)}_{\st y} |  \rho  \right)$.          
The proof that LO follows from the principles, provided in Methods,  consists of three steps:  First, thanks to the Fundamental Sharpness of Measurements, the problem is reduced  to proving that LO holds for probability distributions generated by sharp local measurements.     Then, we observe that, in the case of  sharp measurements, locally orthogonal events correspond to orthogonal effects.     Finally, we use the fact that mutually orthogonal effects  can coexist in a single measurement.  As a corollary, we obtain the bound $\sum_{\st e  \in  \set O}  p  ( \st e)   \le 1$, establishing the validity of LO for all the probability distributions generated by measurements in our theory.   

Like in the case of CE, our principles imply the whole hierarchy of extensions of LO introduced in \cite{Fritz12}. The hierarchy is defined as follows:   the probabilities $p(\st y|\st x)$ satisfy the $L$-th level of the hierarchy if their product $  p(\st y_1|  \st x_1)  \cdots  p(  \st y_L |\st x_L)$ satisfies LO.      
  Now, we can think of the product as being generated by measurements on $N$ copies of the state $\rho$.  In this way, we reduce the problem of proving the $L$-th level of the hierarchy to the problem of proving LO for measurements performed on the state $\rho^{\otimes  L}$.  But we already proved the validity of LO for arbitrary measurements and arbitrary states.    In conclusion, the structure of sharp measurements  implies that LO is satisfied at every possible level.  
  
 A striking consequence of this argument is that the fundamental sharpness of measurements rules out the ultra-strong nonlocality exhibited by PR box correlations, as the latter violate the LO hierarchy \cite{Fritz12}.  In other words, in the world of PR boxes some measurements must be fundamentally noisy.  


\medskip

{\bf Sharp Bell inequalities.}   The request that measurements are ideal at the fundamental level exerts a censorship on the amount of nonlocality that can be detected by experiments.    
To illustrate this fact, we show a number of Bell inequalities where the sharpness of measurements prevents every violation.  We call such inequalities sharp.  

\begin{figure}[floatfix]
\includegraphics[totalheight=90mm]{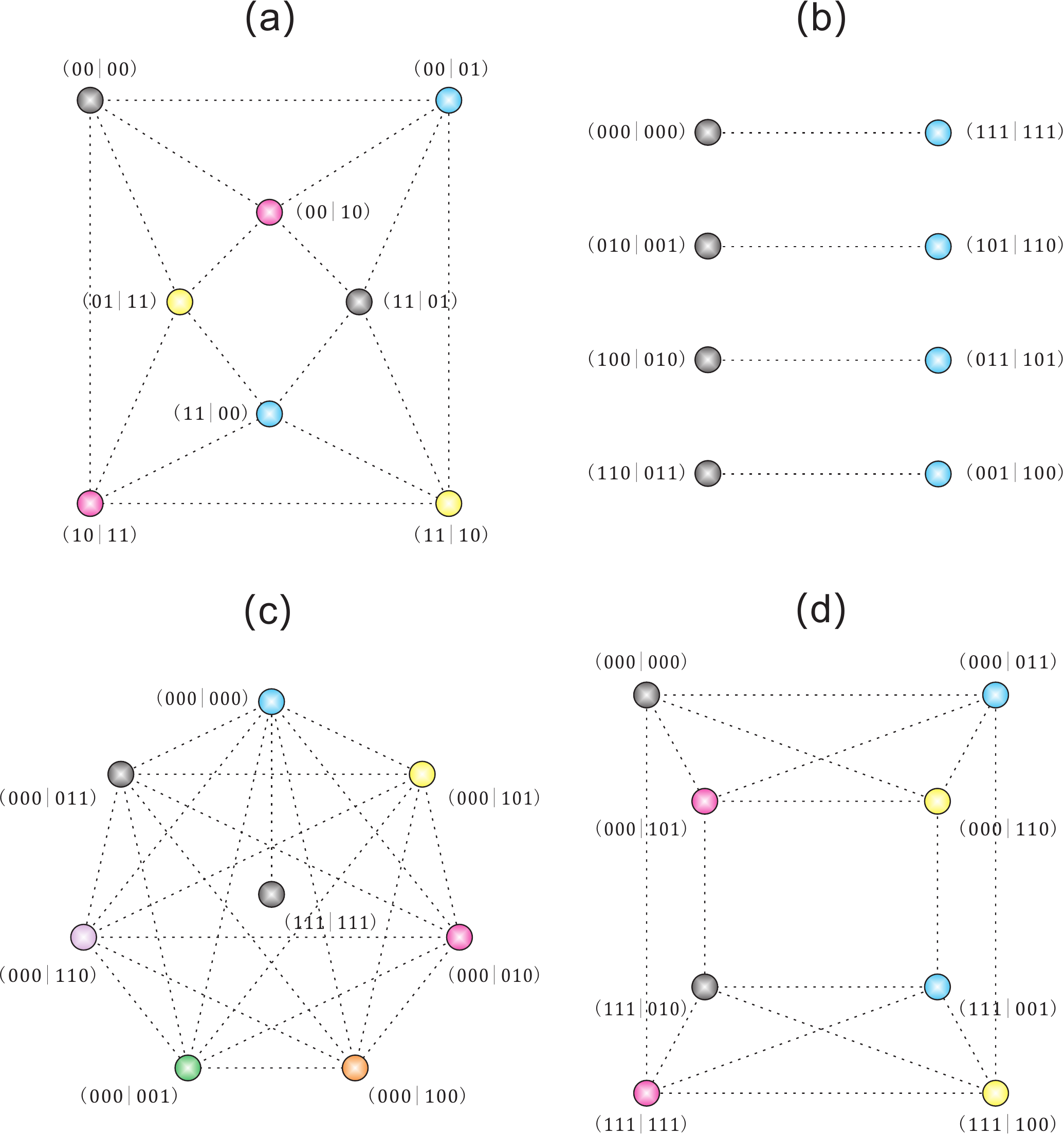}
\caption{{\bf Winning graphs.}  The vertices are coloured  so that two adjacent  vertices have distinct colours, using the minimum number of colours.    ({\bf a})  Winning graph for the CHSH game \cite{chsh}. The player   win   $+1$ if $y_1  \oplus  y_2  = x_1 x_2$  and 0 otherwise.    The graph is not perfect, because the largest clique in the graph has 3 vertices while the number of colours in the graph is 4.  Here classical strategies are not optimal  among the strategies that satisfy LO.       ({\bf b}) Winning graph for Guess Your Neighbor's Input  \cite{Almeida10} in the case of $N=4$ parties. The players win $+1$ if $y_i  =  x_{i+1}$ for every $i$.  The graph is a disjoint union of disconnected cliques and  therefore classical strategies are optimal among all strategies satisfying LO.      ({\bf c})  Winning graph for the game ``Guess the Product" in the case of $N=3$ parties.  The players win  $+1$ if   $y_i  =  x_1x_2 x_3$ for every $i$ and 0 otherwise.   The graph is perfect and therefore the classical strategy is optimal  \cite{acin2012}.         ({\bf d})  Winning graph for the game  ``Guess the Parity" in the case of $N=3$ parties.  The players win  + 1 if $y_i  =  x_1 \oplus  x_2 \oplus x_3$ for every $i$ and 0 otherwise.  Also in this case, the graph is perfect and the classical strategy is optimal. 
}
\label{fig:graphs}
\end{figure}

Consider a game played by $N$ non-communicating parties and a referee, who sends to party $i$ an input $x_i$ and receives back an output $y_i$.   The referee chooses the input string   $\st x$ at random with probability $q(\st x)$ and assigns a payoff $\omega  (\st x,\st y)$ to the players, assumed without loss of generality to be nonnegative for every $(\st x,\st y)$.  The expected payoff obtained by the players is given by $  \omega  =   \sum_{\st x,\st y}    q(\st x)  \omega  (\st x,\st y)   p(\st y|\st
 x)$,      where $p(\st y|\st x)$ is the probability distribution describing their strategy.       
For a given game, the maximum payoff that can be achieved by classical strategies---call it $\omega_c$---defines a Bell inequality,  $\omega  \le \omega_c$. The  game can be associated with a graph  $\set G$, here called the \emph{winning graph}, by choosing as vertices the events  $(\st x, \st y)$ such that  $q(\st x)  \omega  (\st x,\st y)  \not = 0$ and placing an edge between two events  $\st e$ and $\st e'$  if they are not locally orthogonal, as illustrated in Figure  \ref{fig:graphs}. 
In this picture, the maximum payoff  achieved by classical strategies is  
\begin{equation}
\omega_c =  \max_{\set C  \subseteq \set G}   \sum_{  (\st x,\st y)\in\set C }    q(\st x) ~ \omega  (\st x,\st y)  \, ,     
\end{equation}  
where $\set C$ is a clique, \ie  a subset of $\set G$ with the property that every two vertices in $C$ are connected \cite{Fritz12}. 

A first class of games leading to sharp Bell inequalities  is the class of games with a graph $\set G$ that is the disjoint union of mutually disconnected cliques  $\set C_k,    \,  k\in  \{1,\dots, K\}$.     This class contains the game  Guess Your Neighbor's Input \cite{Almeida10}  and the maximally difficult Distributed Guessing Problems of Ref. \cite{Fritz12}.  
In addition, it contains other games such as, for even $N$,  the ``Guess the Parity" game where each player is asked to guess the  parity of the input string  $\st x$.  
For all these games, LO implies that the classical payoff is an upper bound.   Indeed,   picking    for every  $k$  the event  $\st e_k  \in\set C_k$  that has maximum probability, the payoff can be bounded as 
\begin{eqnarray*}
\omega   &  = & \sum_{k=1}^K        \sum_{(\st x ,\st y)\in \set C_k }    q(\st x) ~   \omega  (\st x , \st y)    ~ p\left (\st y|   \st x \right)  \\
 & \le&  \sum_{k  = 1}^K       p\left ( \st e_k \right)          \left[ \sum_{(\st x,\st y)  \in \set   C_k}       q(\st x)   ~ \omega (\st x, \st y  )     \right]   
\end{eqnarray*}
and since the events $\{  \st e_k\}_{k=1}^K$ are locally orthogonal by construction,  one has  $\sum_k  p(\st e_k)  \le 1$ and, therefore,  $\omega   \le   \max_k    q  \left(  \set   C_k \right)  \equiv  \omega_c $.  In conclusion, every game  where the wining graph is a disjoint union of disconnected cliques defines a  Bell inequality,  $\omega  \le \omega_c$  that cannot be violated by any theory satisfying LO, and, in particular, by any theory where measurements are fundamentally repeatable and with minimal disturbance. 
Using a result of Ref. \cite{acin2012},  the proof that LO cuts the payoff down to its classical value can be extended  to a larger class of games, defined by the property that the winning graph $\set G$ is a perfect  graph  \cite{BergeGraphs} (recall that a graph is perfect if, for every subset $\set S$ of its vertices, the number of colours needed to label adjacent vertices in $\set S$ with different colours is equal to the number of vertices in the largest clique $\set C \subseteq \set S$).  Examples of such games are the  ``Guess the Parity" game and the ``Guess the Product" game where the players are win if they guess the product of their inputs.  

\section{Discussion}   


Our results are derived in a minimal framework,  which avoids some assumptions  commonly made   in  GPTs. In particular,  our arguments  do not invoke local tomography \cite{hardy01,mauro2006,barrett2007}, but only the requirement that sharp measurements are local.   This requirement is strictly weaker: for example,  it is satisfied by quantum theory on real  Hilbert spaces, where local tomography fails.   The validity of our results does not even require   that the states of a given physical system  form a convex set.  Thanks to this feature, the results apply  also to  non-convex theories, like Spekkens' toy theory \cite{SpekkensToy}. Interestingly, probabilities themselves do not play a crucial role in our arguments and it is quite straightforward to extend the   results to theories that  only specify which outcomes are possible, impossible or certain, without specifying their probabilities,  such as Schumacher's and Westmoreland's toy theory \cite{ModalQT}.   

Since sharp measurements play  a central role in quantum mechanics,  it is not surprising that they have been the object of extensive investigation since the early days \cite{vonNeumann,Luders50,Dirac}.   
Later, Holevo proposed a purely statistical definition of  sharp measurement, which does not refer to post-measurement states  \cite{HolevoObservable}. Although in the quantum case Holevo's definition  reduces to that of projective measurement, in general theories it is inequivalent to ours, and it is not clear how one could use it to derive features like LO and CE.    Different notions of ideal measurements were put forward by Piron \cite{PironBook,PironIdeal} and Beltrametti-Cassinelli \cite{Beltrametti} in the framework of quantum logic.  In general, they differ form our definition in the way the condition of minimal disturbance is defined.  Most recently,  measurement disturbance  came back  to play  a key role in the search for basic physical principles, as shown \eg by the No Disturbance  Without Information principle of Ref. \cite{Pfister} and by the No Disturbance principle of Ref. \cite{kucaka}.    

Our work joined the insights from two different approaches to the foundations of quantum mechanics: the characterization of quantum correlations \cite{pr95,Dam05, brassard2006,linden2007,Paw09,Navascues09,Fritz12}  and the study of general probabilistic theories  \cite{hardy01,chiribella11,hardy11,masanes,Brukner,masanes12}. 
  Although these two approaches have developed on separate tracks so far,   they share the same fundamental goal: understanding which picture of Nature lies behind the laws of quantum mechanics and guiding our intuition towards the formulation of new  physical  theories and new information-processing protocols.      Our results demonstrate that the interaction between the two approaches can be highly beneficial for both.    Here  LO and CE stimulated the search for new  principles in the GPT  framework, leading to  a compelling picture of nature where measurements are repeatable and cause minimal disturbance  at the fundamental level. It is an open question whether other information-theoretic features of quantum correlations, such as Non-Trivial Communication Complexity, No-Advantage in Nonlocal Computation, and Information Causality  can be derived in this picture.    
  Furthermore,   the idea that a noisy physical process can be reduced to an ideal process at the fundamental level reminds immediately of another quantum feature: Purification \cite{chiribella10}. Operationally, Purification is the property that every mixed state can be generated  from a pure state of a composite system by discarding one component.    This principle implies directly entanglement    and is at the core of the reconstruction of quantum theory of Ref.   \cite{chiribella11}.  
  Our result suggests the possibility that purification and sharpness could be sufficient to single out quantum theory among all possible GPTs. 
 In terms of quantum correlations, this would lead to the tantalizingly simple picture ``Purification brings nonlocality in, sharpness cuts it down".    Going even further, it is intriguing to envisage that purification and sharpness could be viewed as two sides of the same medal by imposing that physical theories must satisfy  a   requirement of time symmetry, similarly to what was done in quantum theory by Aharonov, Bergmann and Lebowitz \cite{ABL,APT}.

\section{Methods}

{\bf Characterization of sharp instruments.} The starting point of our results is the observation that   an  instrument $\{\map M_x\}$ is sharp if and only if
\begin{equation}\label{alte}
(  r_{xy}  |    \map M_x  =   (r_{xy}|    \qquad  \forall x\in \set  X  \, , \forall y \in \set Y     
\end{equation} 
for every measurement $\st r  =  \{r_{xy}\}_{(x,y)\in\set X\times \set Y}$ that refines $\{m_x\}$, \ie  $ \sum_{y\in\set Y}  r_{xy}  =  m_x$ for every $x$.     
 Let us see why Eq. (\ref{alte}) is equivalent to sharpness.   First, suppose that Eq. (\ref{alte}) holds. Clearly, this implies that $\st m$ is repeatable, as one can see by summing over $y$.  Moreover,  Eq. (\ref{alte})  implies that $\st m$ is a minimal disturbance measurement. Indeed, take a generic measurement  $\st n$ that is compatible with $\st m$.   By definition, this means that  there exists a joint measurement $\st r  = \{  r_{xy}\}$ such that $\sum_x r_{xy}  = n_y$ for every $y$ and $\sum_y  r_{xy}  = m_x$ for every $x$.   
 We then obtain  
\begin{align}
\nonumber (n_y|   \map M  &  =  
 \sum_{x}  (r_{xy}|    \map M_{x}   +    (s_y|   \qquad  (s_y| : =    \sum_{x,x':  x\not = x'}  (j_{xy}|    \map M_{x'} \\
 \nonumber &  =   \sum_{x}  (r_{xy}|     +   (s_y|   \\
\label{proof} &  =    (n_{y}|     +      (s_y|  \, ,
\end{align}  
having used Eq. (\ref{alte}) in the second equality.   
 Summing over $y$ and using the normalization of the measurement $\st n$ we obtain the condition $ \sum_y  (s_y|  = 0$, or, equivalently, $\sum_y  (s_y|  \rho)=0$ for every $\rho$.  Since probabilities are non-negative, this implies that each term in the sum vanishes, leading to the relation $s_y=  0$ for every $y$.  Inserting this relation back in Eq. (\ref{proof}) we conclude that $(n_y|  \map M  =  (n_y|$, that is, the instrument does not disturb $\st n$.  Hence,   Eq. (\ref{alte})  implies that $\{\map M_x\}$ is a sharp instrument.         Conversely,  if $\{\map M_x\}$  is a sharp instrument then Eq. (\ref{alte}) must be satisfied.  
  By definition, one has 
  \begin{equation*}
  (  m_x|  =   (e|  \map M_x =  \sum_{x'}  (m_{x'}|  \map M_x  
  \end{equation*}  
  and using the repeatability condition  $(m_x|    = (m_x|  \map M_x  $ one obtains  $\sum_{x'\not = x}   (m_{x'}|\map M_x   =  0$.  
  Again, the fact that probabilities are nonnegative implies  that each term in the sum must vanish, namely 
  $(m_{x'}|\map M_x   =  0 $ for every $x'  \not = x$.
   Now, let $\st r$ be a measurement such that $ \sum_y r_{xy}   =  m_x$.   Since the measurement $\st r$ is compatible with $\st m$, Eq. (\ref{nodist}) implies $ (r_{xy} |  \map M  =  (r_{xy}|$ with $\map M  =  \sum_{x'} \map M_{x'}$.    On the other hand, for $x  \not =  x'$ the condition $( m_x|  \map M_{x'}   =  0$  implies $  (r_{xy}|  \map M_{x'} = 0 $.  Hence, we conclude that $(r_{xy} |  \map M_x  =  (r_{xy} |  \map M   =  (r_{xy}|$ for every $x$ and $y$. \qed 
  
\medskip

{\bf  Joint measurability of orthogonal effects.}     
 The characterization of sharp instruments, combined with the Less Information-More Sharpness principle, leads directly  to the  first key result of our work:  a construction showing that  mutually orthogonal effects can be measured jointly in a single sharp measurement.  Precisely,   if   $m_k$ is orthogonal to $m_l$ for every $k,l  \in\{1,\dots, K\}$, we show that there exists a joint sharp measurement $\st j$ such that $\{m_k\}_{k=1}^K  \subseteq \st j$.   
 
 Let us see how to  construct the  joint measurement.    Let $\st m^{(k)}$ be  the sharp measurement that contains the effect $m_k$.  By coarse graining of $\st m^{(k)}$ one obtains  the binary measurement $\st m^{(k)}  =  \{m_0^{(k)}  ,  m^{(k)}_1\}$, with $m_0^{(k)} : =  m_k$ and $m_1^{ (k) } : =  u-m_k$. By  the Less Information, More Sharpness postulate, $\st m^{(k)}$ is  sharp.    Let  $\{ \map M^{(k)}_0,  \map M_1^{(k)}  \}$ be the corresponding instrument.  Now, since $m_k$ and $m_l$ are orthogonal, $\st m^{(kl)}  = \{m_k, m_l,  e-m_k  -  m_l\}$ must be a valid measurement.     Since $\st m^{(k)}$ is a coarse-graining of  $\st m^{(kl)}$, Eq. (\ref{alte}) gives   
\begin{equation}\label{aa}  (  m_l  |    \map M_1^{(k)}    =   (m_l|\,.
\end{equation}  
 Now, consider the following measurement procedure:  \emph{i)} perform the first instrument   \emph{ii)}  if the outcome is $1$, then perform the second instrument,  \emph{iii)}  for every $k< K $,  if the outcome of the $k$-th instrument is $1$, perform the  $(k+1)$-th instrument.   The resulting instrument, denoted by $\{\map J_k\}_{i=1}^{K+1}$ consists of the transformations
\begin{align*}
\nonumber \map J_{1}   & :=  \map M^{(1)}_{0} \\
\nonumber \map J_{2}   &:=  \map M_0^{(2)}  \map M_1^{(1)}   \\
\nonumber \map J_{3}       &: =  \map M_0^{(3)}   \map M_1^{(2)}    \map M_{1}^{(1)}  \\
 \nonumber &\, ~~\vdots  \\
\map J_{K}   &:=   \map M_{0}^{  (K) } \map M_1^{(K-1)}\cdots    \map M_1^{(1)}  \\
\map J_{K+1}   &:=   \map M_{1}^{  (K) } \map M_1^{(K-1)}\cdots    \map M_1^{(1)}  \, .
\end{align*}
The measurement $\st j  = \{j_k\}_{k=1}^{K+1}$ associated to  the instrument $\{\map J_k\}_{k=1}^{K+1}$ is the desired joint measurement: indeed,  and using Eqs.   (\ref{prob}) and (\ref{aa})  we obtain  $(j_k|  =  (e|  \map  M_0^{(k)}   \map M_1^{(k-1)}  \cdots  \map M_1^{(1)}   =    ( m_k|   \map M_1^{(k-1)}  \cdots  \map M_1^{(1)}     =   (m_k|$ for every $k  \in   \{1,\dots,  K\}$.   In addition, the measurement $\st j$ is sharp.  Indeed, if a measurement $\{r_{kl}\}$  is a refinement of $\st j$, \ie  $  \sum_l  r_{kl}  =  j_k$ for all $k$,  then $\st r$ is also a refinement of the sharp measurement  $\st m^{(k')}$ for every fixed $k'$.   Hence, one has 
\begin{align*}
(  r_{kl}|  \map M_0^{(k)}    &=  (r_{kl}  |   \qquad  \forall k \in  \{1,\dots K\}  \\
(  r_{kl} | \map M_1^{(k')}    & =  (r_{kl}|  \qquad 
\forall k, k' \in  \{1,\dots K\}   ,  \,  k\not= k' \, .  
\end{align*} 
Using this fact and the definition of $\map J_k$ it is immediate to obtain the relation $(  r_{kl}|  \map J_k  =  (r_{kl}|$ for every $k,l$.   Thanks to Eq. (\ref{alte}), this proves that the instrument $\{  \map J_k\}$ is sharp, and so is the corresponding measurement $\st j$.   \qed 

\medskip  

The ability to combine orthogonal effects into a single measurement is a powerful  asset.   As we already observed,  it implies CE at its basic level.  In the following we show that it can be used also to obtain the whole CE  hierarchy.

\medskip

{\bf Orthogonality of product effects.}   
 In a causal theory  the information available at a given moment of time can be used to make decisions about the settings of future experiments, thus allowing for adaptive measurements where the choice of setting for a system   $B$ depends on the outcome of a measurement on system  $A$.   In particular, if $\{m_x\}$ is a sharp measurement on $A$ and $\{  n^{(x)}_y\}_{y\in\set Y}$ is a sharp measurement for every $x$, then $\{m_x\otimes n^{(x)}_y\}$ is a legitimate measurement.  Now, the Locality of Sharp Measurements implies that      $\{m_x\otimes n^{(x_0)}_y\}$ is sharp for every fixed $x_0$.  Since $x_0$ is arbitrary, this means that each effect $m_x\otimes  n^{(x)}_y$ is sharp and that two effects  $m_x\otimes  n^{(x)}_y$  and $m_{x'}\otimes  n^{(x')}_{y'}$ are orthogonal unless $x=x'$ and $y=y'$.


Thanks to this observation,  it is easy to see that  every level of the CE hierarchy is satisfied.  The key is to note that if two strings of outcomes $\st y$ and $\st y'$ are exclusive,  then the corresponding effects $P_{\st y}$  and $P_{\st y'} $ are orthogonal.  This is clear because, by definition,  the effects corresponding to two exclusive strings are of the form $   P_{\st y}   =  m_{y_i}  \otimes  n$  and $P_{\st y'}  =   m_{y_i'} \otimes n'$ where  the effects   $m_{y_i}$ and $m_{y_i'}$ are orthogonal and the effects $n  =  \otimes_{j\not  = i}  m_{y_j}$ and $  n'  =  \otimes_{j\not  = i}  m_{y_j} $ are sharp  thanks to the Locality of Sharp Measurements.    Using our result about product effects, we then have that   $P_{\st y}$ and $P_{\st y'}$ are orthogonal.     
 Now,  a set of mutually exclusive strings  $\set E$ corresponds to a set of mutually orthogonal effects  $\{  P_{\st y}\}_{\st y\in\set E}$.   Since mutually orthogonal effects can be combined into a joint measurement, the probabilities $p_L (\st y)  =   (  P_{\st y}|   \rho^{\otimes L}) $ obey  the bound $\sum_{\st y\in\set E}  p_L(\st y)\le 1$, meaning that the theory satisfies the $L$-th level of the CE hierarchy for arbitrary $L$.

Note that the same argument can be used to prove the validity of LO for the probability distributions generated in a scenario where all parties perform sharp measurements.   In this scenario,  two locally orthogonal events $(\st x,\st y)$ and $(\st x',\st y')$ correspond to two orthogonal effects   $P^{(\st x)}_{\st y}$  and $P^{(\st x')}_{\st y'}$, for exactly the same reason mentioned above.  Hence,  the joint measurability of orthogonal effect implies the bound $  \sum_{(\st x,\st y)  \in\set O} p(\st x|\st y)\le 1$ for every set $\set O$ of locally orthogonal events.    In other words, all the  probability distributions  generated by sharp measurements obey LO.

\medskip 
 
{\bf Reduction to sharp measurements.}  While CE applies only to sharp measurements, LO applies to arbitrary measurements.     This is because  every probability distribution that we can encounter in our theory is a probability distribution generated by sharp measurements.   This fact can be seen as follows:  combining the Fundamental Sharpness with the Locality of Sharp Measurements,   one can show that for  every party $i$ and every measurement $\st m^{(i,x_i)}$, there exists an ancilla   $A_i$, a state of $A_i$, call it $\sigma_i$, and a  sharp measurement $\st M^{(i,x_i)}$  such that $     \left(  m^{(i,x_i)}_{y_i}  |    \rho_i   \right)   =  \left(    M^{(i,x_i)}_{y_i} |   \rho_i\otimes \sigma_i \right)  $ for every $x_i$ and for every $y_i$    
(see the Appendix for the proof).     Now, since all  measurements that party $i$ can perform can be replaced by sharp measurements by adding an ancilla in a fixed state $\sigma_i$, the input-output distribution $p(\st y|\st x)$ generated by arbitrary measurements on the state $\rho$ coincides with the input-output distribution generated by sharp measurements on the state $\rho'  =  \rho  \otimes  \sigma_1\otimes \dots\otimes \sigma_N$.  
In other words, at the level of correlations there is no difference between sharp and non-sharp measurements.  Thanks to this fact, deriving LO for sharp measurements is equivalent to deriving LO   for  arbitrary  measurements.



\begin{acknowledgments}

\subparagraph*{Acknowledgements}  This work is supported by the National Basic Research Program of China (973) 2011CBA00300 (2011CBA00301),  by the National Natural Science Foundation of China through Grants  11350110207, 61033001,  and 61061130540,  by the Foundational Questions Institute through the large grant ``The fundamental principles of information dynamics",  and by the 1000 Youth Fellowship Program of China.  
GC acknowledges  A B Sainz,  A Leverrier, and R Garcia-Patron  for useful discussions and 
acknowledges the  hospitality of the Simons Institute for the Theory of Computing and of Perimeter Institute for Theoretical Physics, where part of this work has been done.  Research at Perimeter Institute for Theoretical Physics is supported in part by the Government of Canada through NSERC and by the Province of Ontario through MRI.
\end{acknowledgments}


%

\begin{appendix}
\section*{Sharp realization of multiple measurements}

Here we show that the Fundamental Sharpness principle implies that for   every party $i$  one can find an ancilla and a state of the ancilla, independent on the setting $x_i$, such that every measurement of party $i$ can be realized as a sharp measurement on the system and the ancilla.  This fact is a consequence of the following: 

\begin{prop}   Let $\{  \st m^{(x)}~,~  x\in\set X\}$  be a finite set of measurements on system $A$ and let $\set Y_x$ be the set of outcomes for measurement $\st m^{(x)}$.   If the Fundamental Sharpness and the Locality of Sharp Instruments  hold, then  there exists an ancilla   $B$, a state  $\sigma$ of the ancilla, and a  sharp measurement $\st M^{(x)}$  such that 
\begin{align*}    
 \left(  m^{(x)}_{y}  |    \rho   \right)   =  \left(    M^{(x)}_{y} |   \rho\otimes \sigma \right)      \qquad \forall x\in \set  X \, , \forall y\in\set  Y_x \, 
 \end{align*}
 for every state $\rho$ of system $A$.  
\end{prop}

{\bf Proof.}  For every setting $x$, the Fundamental Sharpness principle ensures that  there exists an ancilla  $A_{x}$, a state $\sigma_{x}$ of the ancilla,   and  a sharp measurement   $\st S^{(x)}$ such that, for every state $\rho$, one has 
\begin{align*}  \left(m^{(x)}_{y}  |   \rho\right) =  \left(   S^{( x)}_{y}  |    \rho \otimes \sigma_{x} \right)  \qquad \forall y\in\set Y_x,   
\end{align*} 
Now, take as ancilla the  composite system $B  :  =  \bigotimes_{x}   B_x  $,    define the state $\sigma :  =   \bigotimes_{x}     \sigma_{x}$,    and let   $\st M^{(x)} $  be the measurement with effects $  M_{y}^{(x)} :  =    S^{(x)}_{y}\otimes u_{\lnot x}$, where $ u_{\lnot x}$ is the unit effect on all ancillas $A_{x'}$ with $x' \not  = x$. Now, $\st M^{(x)}$ is the product of the sharp measurement $\st S^{(x)}$ with the trivial measurement, which, by definition, is also sharp.  Hence, by the Locality of Sharp Measurements,  $\st M^{(x)}$ is sharp.   By construction,  one has  
\begin{align*}
 \left(M^{(x)\prime}_{y}  |   \rho\otimes \sigma\right)   &  =  \left(   S^{(x)}_{y}  |    \rho \otimes \sigma \right)  \\
  &  =   \left(m^{(x)}_{y}  |   \rho_i\right)  
 \end{align*}
for every setting $x$, every outcome $y$, and every state $\rho$.   \qed 

\end{appendix}

\end{document}